# Human-in-the-Loop Simulation for Real-Time Exploration of HVAC Demand Flexibility

Xinlei Zhou*, Han Du, Emily W. Yap, Wanbin Dou, Mingyang Huang, Zhenjun Ma*

Sustainable Buildings Research Centre, University of Wollongong, 2522, Australia

**Abstract**

The increasing integration of renewable energy into the power grid has highlighted the critical importance of demand-side flexibility. Among flexible loads, heating, ventilation, and air-conditioning (HVAC) systems are particularly significant due to their high energy consumption and controllability. This study presents the development of an interactive simulation platform that integrates a high-fidelity simulation engine with a user-facing dashboard, specifically designed to explore and demonstrate the demand flexibility capacity of HVAC systems. Unlike conventional simulations, where users are passive observers of simulation results with no ability to intervene in the embedded control during the simulation, this platform transforms them into active participants. Users can override system default control settings, such as zone temperature setpoints and HVAC schedules, at any point during the simulation runtime to implement demand response strategies of their choice. This human-in-the-loop capability enables real-time interaction and allows users to observe the immediate impact of their actions, emulating the practical decision-making process of a building or system operator. By exploring different demand flexibility scenarios and system behaviour in a manner that reflects real-world operation, users gain a deeper understanding of demand flexibility and their impacts. This interactive experience builds confidence and supports more informed decision-making in the practical adoption of demand-side flexibility. This paper presents the architecture of the simulation platform, user-oriented dashboard design, and user case showcase. The introduced human-in-the-loop simulation paradigm offers a more intuitive and interactive means of engaging with grid-interactive building operations, extending beyond HVAC demand flexibility exploration.

**1. Introduction**

As energy systems transition toward low-carbon futures, the growing share of variable renewable generation has made grid flexibility a critical requirement. While supply-side measures remain important, increasing attention is now also being directed to the demand side, where real-time load adjustment can play a vital role in maintaining grid stability and reliability [1]. Buildings, which account for approximately 30% of global final energy consumption, have emerged as promising contributors to demand-side flexibility, particularly through demand response (DR) strategies [2].

Among building systems, heating, ventilation, and air-conditioning (HVAC) systems represent a primary opportunity for DR due to their substantial energy use, controllability, and inherent thermal inertia. These characteristics allow for temporary load reductions or shifts with minimal disruption to occupant comfort [3, 4]. By adjusting HVAC operation during peak demand periods or renewable generation fluctuations, buildings can help alleviate grid stress and enhance the integration of clean energy resources [5]. As a result, HVAC systems are central to the vision of grid-interactive efficient buildings and the broader goal of enabling responsive, resilient energy infrastructure.

Despite the technical potential, the practical implementation of HVAC-based DR strategies remains limited largely due to concerns about their impact on control reliability, energy efficiency, and occupant thermal comfort [6]. In response, simulation has become an essential tool for developing and testing DR strategies in a risk-free environment, enabling researchers to explore different scenarios before real-world application. However, traditional simulation platforms, such as EnergyPlus [7, 8], TRNSYS [9, 10], and Dymola [11, 12], are primarily designed for offline analysis. These tools treat simulation models as static testbeds, requiring users to configure scenarios prior to runtime and to access outputs only after the simulation has completed or as they unfold. This setup positions users as passive observers rather than active participants, limiting their ability to intervene in real time or explore control decisions in ways mirroring real-world building operations. Without the opportunity to interact directly with the simulation, users are constrained in understanding system behaviour, the characteristics of demand flexibility, and the practical trade-offs of DR strategies. This lack of engagement and inability to replicate real-world operational scenarios can ultimately weaken user trust and hinder the adoption of DR measures in practice.

To address these limitations, this project introduces a paradigm shift in simulation practice, from passive demonstration to active exploration. We present an interactive, human-in-the-loop (HiL) simulation platform that embeds the user directly into the control loop, enabling real-time adjustment of HVAC control parameters such as temperature setpoints and system on/off scheduling. The platform allows users to engage with the simulation as if they were building operators, exploring the impacts of control decisions on power consumption, thermal comfort, and system dynamics as the simulation runs. This transforms simulation into a virtual operating environment for the exploration and demonstration of HVAC demand flexibility. As buildings continue to evolve into flexible, responsive assets within future smart grids, such HiL simulation platforms will be critical to advancing both technological innovation and human-centred control in demand-responsive HVAC systems.

**2. Methodology**

The HiL simulation platform is designed not only to replicate the real-time operation of buildings and their HVAC systems under varying operating conditions, but also to support user-initiated control and interactions at any time during the simulation. The platform integrates high-fidelity models of both the building envelope and the HVAC plant, developed in EnergyPlus and Dymola, respectively. To ensure interoperability, both models are exported as Functional Mock-up Units (FMUs) [13] and co-simulated in Python leveraging the PyFMI package [14]. Python manages real-time data exchange, synchronises the simulation through a fixed time-stepped loop, also implements supervisory control strategies to govern system operation.

The co-simulation engine functions as the backend, while a browser-based dashboard developed using the Dash framework by Plotly [15] serves as the frontend, as shown in Fig. 1. The dashboard enables users to configure simulation parameters prior to execution and to adjust control settings, such as temperature setpoints and HVAC schedules, during runtime. Real-time plots display indoor temperatures, HVAC power consumption, and setpoint trajectories, allowing immediate visual feedback. The platform also includes a baseline scenario representing business-as-usual system performance, which can be directly compared to controlled scenarios. An energy comparison panel highlights differences in cumulative power consumption, offering clear insights into the effects of user interventions. All user interactions are processed through Dash callbacks and a central state dictionary, ensuring low-latency updates and smooth communication between the dashboard and the backend. The integration of detailed building and HVAC models, an interactive dashboard, and Python-based synchronisation creates a versatile tool for both research and stakeholder engagement in exploring HVAC demand flexibility.

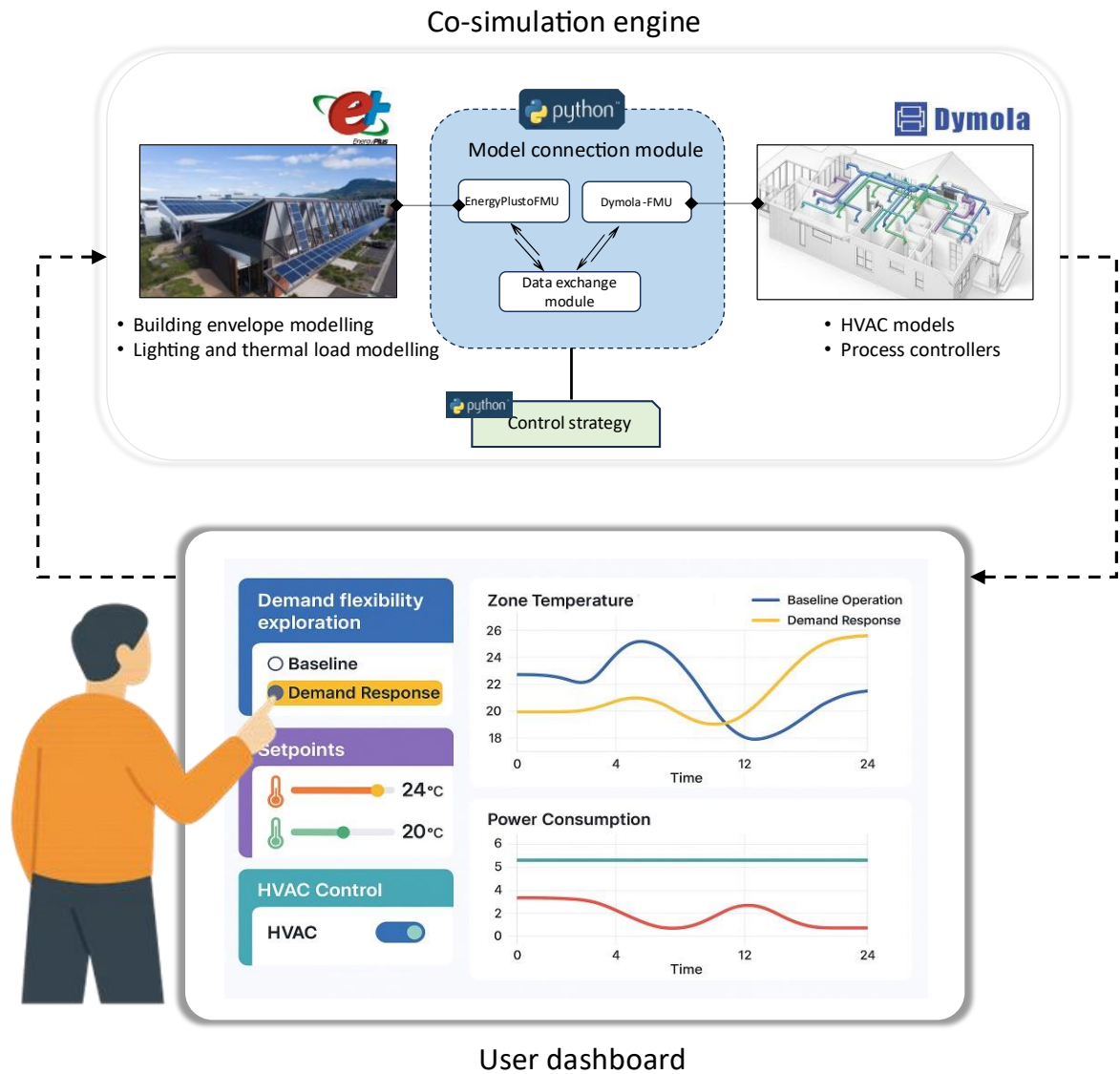

Fig. 1. Architecture of the HiL simulation platform.

## 3. Case study building for platform development

The HiL simulation platform was developed based on the Sustainable Buildings Research Centre (SBRC) at the University of Wollongong as the case study, as shown in Fig. 2. The building comprises two distinct sections: a naturally ventilated northern wing that houses experimental facilities, and a two-storey office building (southern wing), which is served by an active HVAC system. This study focuses exclusively on simulating the HVAC operation in the southern wing.

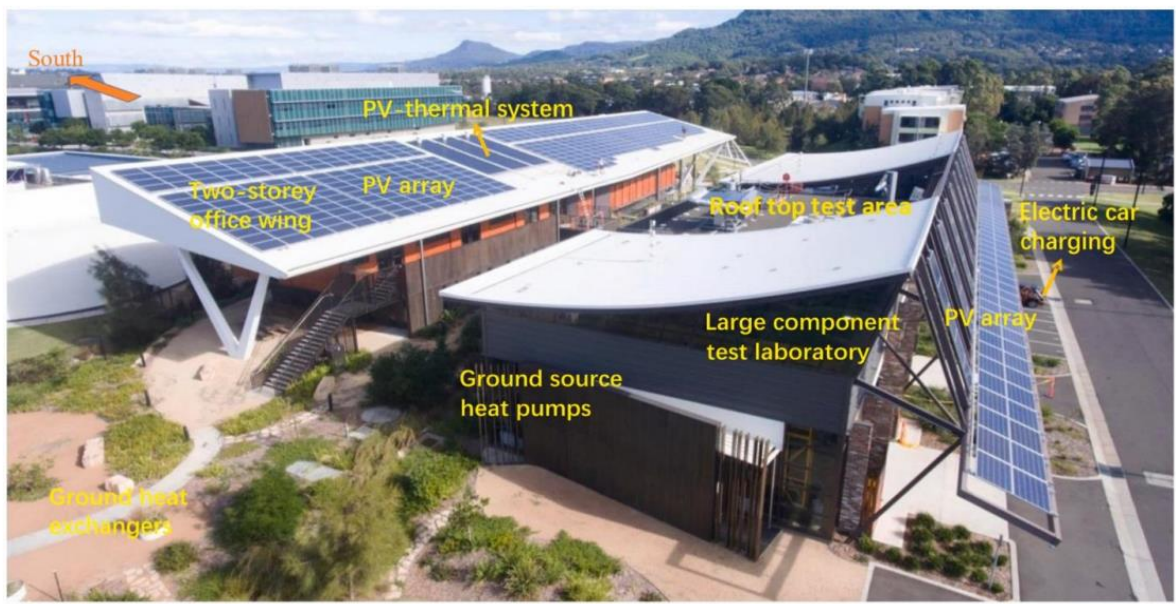

Fig. 2. Bird's eye view of the SBRC [16].

The HVAC system in the SBRC building (Fig. 3) consists of three heat pumps: one air-source (ASHP) and two ground-source (GSHPs). These supply either chilled or heated water to two air handling units, which deliver hot or cold air to the building's zones through variable air volume (VAV) boxes. Each VAV box adjusts airflow using PI controllers to maintain its zone's temperature setpoint. Similarly, PI controllers regulate the control valves on the heating and cooling coils, which manage water flow from the load-side pumps.

This multi-source setup of the HVAC plant allows the system to switch between the ASHP and GSHPs depending on demand. The GSHPs are used for base load coverage, while the ASHP is activated during peak demand periods. Following the methodology described in Section 2, the SBRC building was modelled in an HiL simulation environment, named the SBRC HVAC Virtual Flexibility Lab (SBRC-HV-FlexLab).

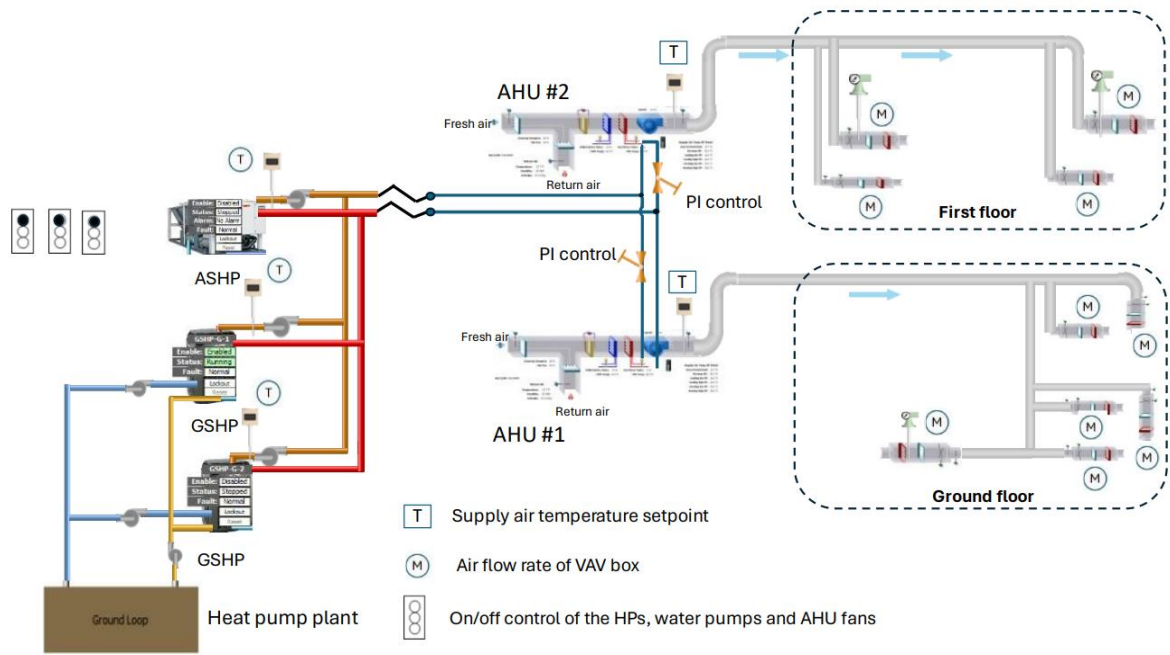

Fig. 3. Schematic of the HVAC system.

## 4. Demonstration of demand flexibility optimisation on the HiL simulation platform

To demonstrate the performance and effectiveness of the SBRC-HV-FlexLab, different interactive demand response scenarios were conducted over 24 hours under representative cooling weather conditions. To benchmark the impact of DR strategies on system performance, the simulation enables the real-time comparison between the baseline scenario and DR scenario, which are designed as follows:

1) <u>Baseline operation:</u> In the default operation scenario, the HVAC system operates from 7:00 to 18:00, with a cooling setpoint of 24 °C and a heating setpoint of 19 °C.

Simulation results under these conditions are used as the baseline for benchmarking the controlled scenarios.

2) <u>Demand response:</u> In the current implementation, DR is achieved by modifying global cooling temperature setpoints or adjusting HVAC operating hours. To facilitate user interaction, setpoint adjustments are represented using discrete control modes: -2, -1, -0.5, 0, 0.5, 1, and 2. These modes correspond to deviations from the baseline cooling setpoint, where 0 indicates no change, -0.5, -1 and -2 represent reductions of 0.5°C, 1 °C and 2 °C, respectively, and 0.5, 1 and 2 represent increases of 0.5°C, 1 °C and 2 °C, respectively. Alternatively, users can manually set other desired values by editing the input boxes to override the zone temperature setpoint or HVAC operation hours. The HVAC operation start and end times can also be overridden during the simulation.

Two controlled scenarios, including a load-shifting scenario and a peak shaving scenario, are presented to demonstrate the performance of the SBRC-HV-FlexLab platform.

4.1 Load shifting scenario

In the load-shifting scenario, the HVAC system's start time is moved one hour earlier, changed from 7:00 (420 min after midnight) to 6:00 (360 min after midnight), and the cooling setpoint is reduced by 2 °C between 6:00 and 7:00 to enable precooling. The setpoint is restored to the baseline value at 7:00, marking the start of the normal operation period. The dashboard illustrates the results of the load-shifting scenario (Fig. 4). It can be observed through the live data that, in this setup, the HVAC system is operated more intensively during the user-initiated DR event, lowering the indoor temperature below the default setpoint during non-occupancy hours. This deliberate precooling phase reduces cooling demand during the typical start-up period of occupancy hours.

The energy comparison chart confirms the effectiveness of this strategy: despite a higher energy use during the pre-cooling hours, the controlled scenario achieves a 4.2% energy saving during the non-DR period compared to the baseline. It is noteworthy that the DR period is defined as the time during which the system operates differently from the baseline, such as through setpoint adjustments or changes to HVAC operating hours. A supplementary video is provided at *https://github.com/XLZo/HiL-Simulation-Demo* to illustrate the live simulation and user-initiated control overrides conducted during the experiment. The results demonstrate the capability of the SBRC-HV-FlexLab to simulate and visualise demand flexibility in response to real-time user inputs, confirming its value for exploring DR strategies in building HVAC systems.

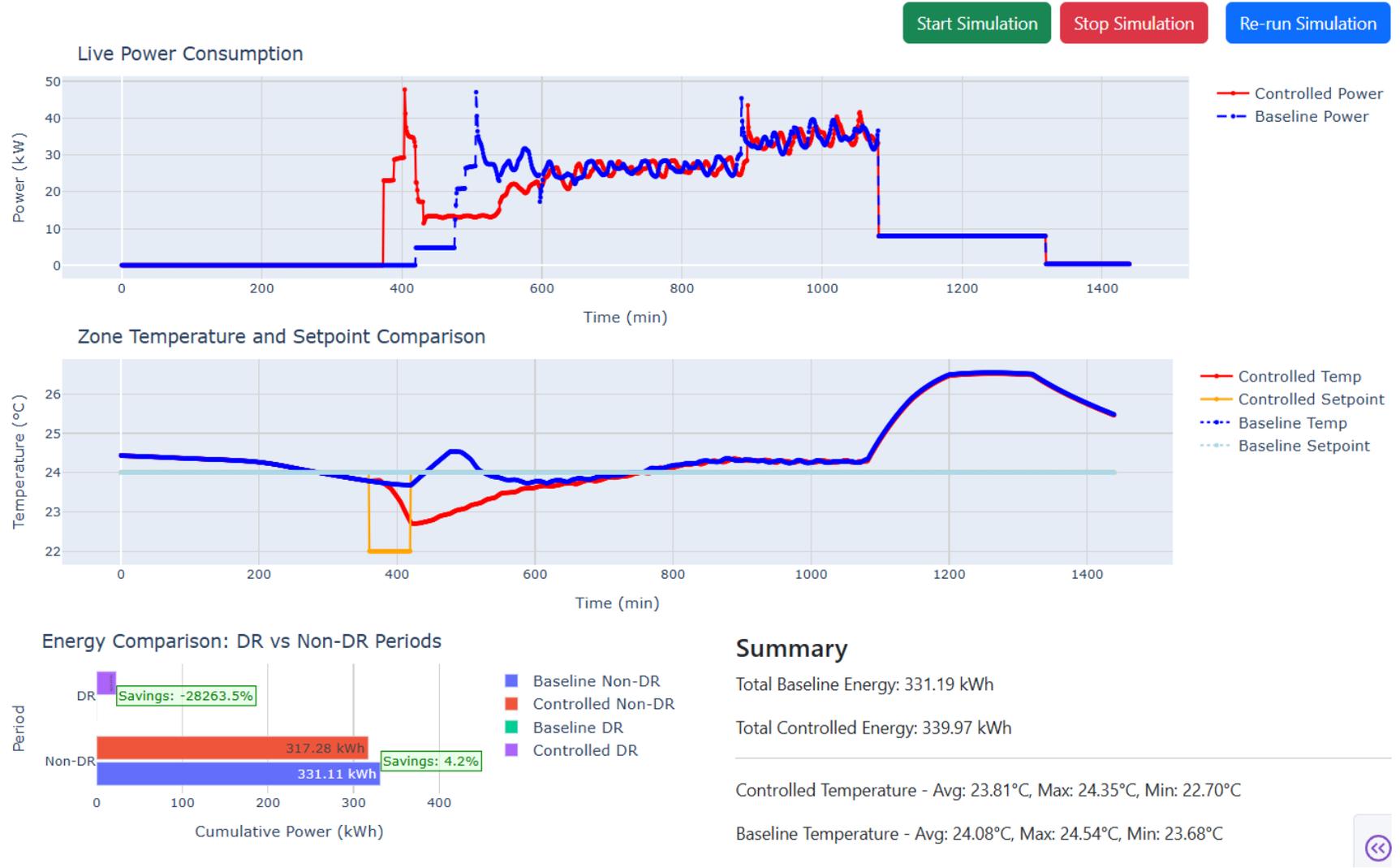

Fig. 4. Results of the load shifting scenario.

4.2 Load shedding scenario

A load shedding scenario was conducted in the SBRC-HV-FlexLab, where the zone temperature setpoint was increased by 1 °C at 12:00 (720 min after midnight) and restored to the baseline at 14:00 (840 min after midnight) to temporarily reduce power consumption during peak demand hours. The results are presented in Fig. 5. During the DR period, the HVAC system in the controlled scenario significantly reduced its operation, as evidenced by a sharp drop in power usage on the live power graph. Notably, a rebound effect occurred immediately after the DR event, reflected by a brief increase in power consumption. Throughout the DR period, the indoor temperature drifted upward but remained within the adjusted setpoint, returning to baseline levels afterwards and maintaining thermal comfort.

The energy comparison chart in the bottom left highlights the effectiveness of this strategy. During the DR period, the controlled scenario achieved a 17.4% reduction in energy consumption compared to the baseline. Although a slight increase in energy use was observed after the DR period, the total energy consumption over the 24-hour simulation was lower in the controlled case (327.55 kWh) than in the baseline (331.19 kWh), confirming the effectiveness of this passive load shedding approach.

These results enable users to quickly understand the impact of DR operations, gain insight into system behaviour, and become more aware of the potential for demand flexibility. A supplementary video is available at *https://github.com/XLZo/HiL-Simulation-Demo*, illustrating the live simulation process conducted during the experiment.

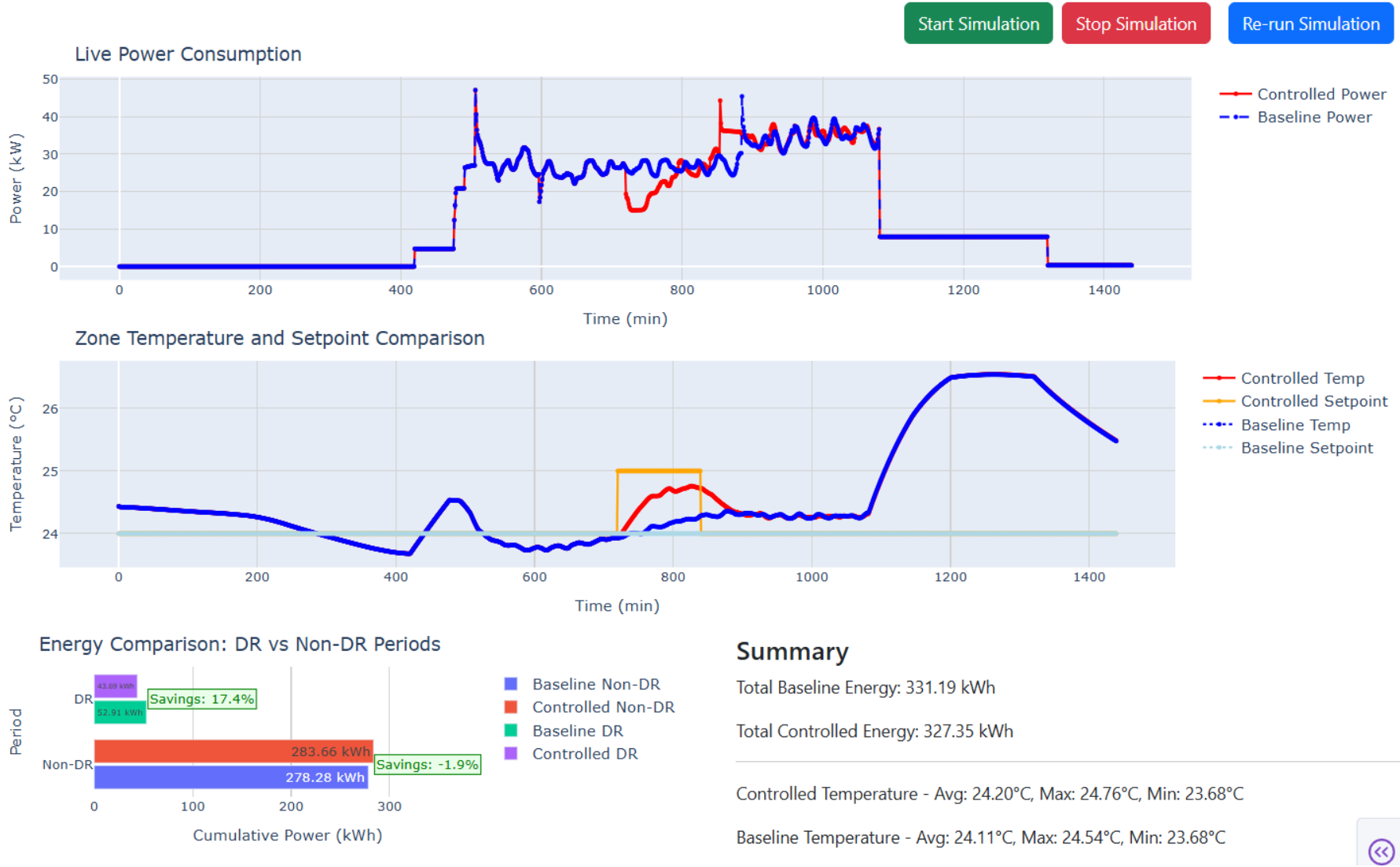

Fig. 5. Results of the load shedding scenario.

4.3 Applications and broader impacts

Pilot testing with a small group of researchers and engineers demonstrated strong usability of the platform. Participants found the live experimentation and visualisation capacities to be highly effective and engaging, and saw potential for enhancing understanding and awareness of demand flexibility. The platform's FMU-based backend architecture also supports the integration of new control strategies from rule-based to AI-driven approaches, transforming it into a flexible sandbox for designing and testing a wide range of demand-side management strategies. More details of the interfaces to integrate advanced controllers will be detailed in a separate publication.

Overall, by enabling HiL simulation, this platform serves as a critical bridge between high-fidelity simulation and practical building operation. As the need for grid-responsive, energy-efficient buildings grows, it stands out as a valuable enabler for innovation in demand flexibility and smart building management.

## 5. Conclusion

In this study, we presented a human-in-the-loop simulation platform that integrates a user-interactive dashboard with a high-fidelity building HVAC model to support demand flexibility exploration. Moving beyond traditional static simulations, the platform enables real-time, user-initiated experimentation to explore and evaluate demand response strategies, assessing their impacts on HVAC system performance in a manner that more closely mirrors real-world building operations. Beyond its immediate value for education and stakeholder engagement, the platform reflects a broader shift in how building control systems are modelled, tested, and deployed. Its interactive and transparent design lays the groundwork for a new generation of simulation tools well-suited to the digital and occupant-centric future of energy systems. As the demand for resilient, energy-efficient buildings continues to grow, such platforms will play a pivotal role in translating innovative control strategies into practical applications in the built environment. A detailed description of the co-simulation engine of the platform will be included in a forthcoming publication. Future work will integrate additional real-time data sources to enable even more informed and effective demand flexibility exploration.